# Modelling of Statistical Low-Frequency Noise of Deep-Submicron MOSFETs

*Gilson I Wirth[1,2], Jeongwook Koh[3], Roberto da Silva[2], Roland Thewes[3], and Ralf Brederlow[3]*

[1]**State Univ. of Rio Grande do Sul-UERGS, Estrada de Santa Maria 2300, 92500 Guaíba RS, Brazil**

[2]Federal Univ. RGS-UFRGS, Porto Alegre, RS, Brazil

[3]Infineon Technologies, Corporate Research, Munich, Germany

**e-mail: gilson-wirth@uergs.edu.br   phone: +55 51 9915-3906   fax: +55 51 3316-7308**

## Abstract

*The low-frequency noise (LF-noise) of deep submicron MOSFETs is experimentally studied with special emphasis on yield relevant parameter scattering. A novel modeling approach is developed which includes detailed consideration of statistical effects. The model is based on device physics-based parameters which cause statistical fluctuations in LF-noise behavior of individual devices. It can easily be implemented in a compact model for use in circuit simulation tools.*

## 1 - Introduction

Low-frequency noise is a performance limiting factor in many of today's CMOS analog and RF circuits. Aiming for robust circuit design, it is essential to develop a detailed understanding of the devices' noise behavior.

Recent works show that the LF-noise performance of modern small area MOS devices is dominated by Random Telegraph Signal (RTS) fluctuations [1, 2]. Their origin is the capture and subsequent emission of charge carriers at discrete trap levels near the Si–SiO$_2$ interface. Noise performance may strongly vary between different devices on one chip, and moreover even between different operation points of a single device. Although LF-Noise has deserved great attention, today, detailed statistical models are not available. Due to the even yield restricting effect of low frequency noise in many applications (e.g. in wireless transceiver designs) the need to statistically model the noise behavior increases, in particular for future analog and RF products. This work is aiming for provision of a comprehensive understanding of the fluctuations of LF-Noise performance of deep submicron devices.

The paper is organized as follows: After a brief discussion concerning the basics of RTS noise, a statistical modeling approach is presented based on the physical origin of the LF-noise in section 3. There, the dependence of noise performance on device geometry and operation point is studied in detail. The model is compared to experimental data from 3 different technology nodes, $0.25\mu m$ ($t_{ox}$=5nm), $0.13\mu m$ ($t_{ox}$=2.2nm), and $0.09\mu m$ ($t_{ox}$=1.6nm) in section 4. Finally, in section 5, the paper is concluded.

## 2 RTS- and 1/f-Noise

In this paragraph the average low frequency noise of individual MOSFETs is briefly reviewed, and some important parameter for the statistical evaluations following in chapter 3 are introduced. The low frequency noise of small area devices shows Lorentzian-like spectra as shown in Fig. 1. Strong fluctuations are observed for the spectra of different devices with same geometries and from the same chip. In Fig. 2, the strong dependence of RTS noise on the bias point is shown. This behavior poses great challenges to design for high yield of minimal area low noise analog and RF circuits in advanced CMOS technologies. Deviations of orders of magnitude are observed between individual devices and at different operating points of a single device (cf. Figs. 1 and 2).

Noise spectra of today's small area devices are believed to be dominated at least for a certain frequency band by RTS from single trap states. This assumption has been experimentally confirmed by time domain RTS measurements from several groups [2, 4].

As basis for the statistical modeling, the physics behind low frequency noise phenomena are discussed here with special emphasis on their microscopic nature. Traps located in the gate oxide near the interface to the silicon capture and reemit some of the carriers responsible for the drain current flowing between source and drain of the device [8]. The impact of the variation of the charging state of these traps on drain current has a similar effect as a fluctuation of the gate voltage. Therefore, an equivalent gate voltage fluctuation is frequently used to derive a simple equation for MOS low frequency noise. In this paper, all experimental results are presented as equivalent gate voltage fluctuations.

To derive an approximation for the low frequency noise we start with a simple equation for the drain current [12]:

$$I_d = q \cdot W \cdot \mu(y) \cdot N_c(y) \cdot \frac{\partial V(y)}{\partial y} \qquad (1a)$$

and $I_d \approx g_m(V_{g,eff}) \cdot V_{g,eff}$ $\qquad (1b)$

Here, $q$ is the elementary charge, $W$ the device width, $\mu(y)$ the mobility at location $y$ in the channel, $g_m$ the transconductance, $N_c(y)$ the number of free carriers per infinitesimal length, $\partial V(y)/\partial y$ the local lateral electrical field in the transistor, and $V_{g,eff}$ is the effective gate-voltage, defined by the difference between gate-voltage and threshold voltage. $N_c(y)$ is given by the following equation [14]:

$$N_c(y) \cong \left( V_{g,eff} - V(y) \right) \cdot C_{ox}/q \qquad (2)$$

Here, $V(y)$ is the difference between channel potential at location $y$ and source-voltage, $C_{ox}$ the gate capacitance per area. Eq. (2) is normally used for transistor operating conditions in the linear mode [14], but it is also valid for transistor operating points under saturation conditions to describe the current in the channel region between source and pinch-off point. Since this is the by far most relevant region for 1/f-noise considerations it is suitable for our purposes in all cases.

For sub-$V_{th}$ operating points, the approximation

$$V_{g,eff} = 2 \cdot n \cdot kT \frac{\ln(1 + \exp(qV'_{g,eff}/2nkT))}{1 + C_{ox} \cdot \sqrt{\dfrac{2\Psi_s}{q\varepsilon_{Si}N_{ch}} \cdot \exp(qV'_{g,eff}/2nkT)}} \qquad (3)$$

is used for the effective gate voltage [15]. The term $kT$ is the thermal energy, $V'_{g,eff}$ is the effective gate voltage as applied to the terminals of the device. This equation is e.g. used in the BSIM models. In strong inversion $V'_{g,eff}$ is similar to $V_{g,eff}$. For sub-$V_{th}$, the formula allows a monotonic enhancement of the classical current formulation (eq. 1). A discussion of other parameters in eq. (3), which are not important in this context, is given in [15]. Eqs. (2) and (3) give us a drain current formula for deriving the standard deviation of the drain current (in other words: the current noise) due to trap influence.

The influence of the traps on the drain current is twofold. On the one hand, the occupation of a trap changes the number of free carriers in the channel, on the other hand, a charged trap state has a strong influence on the local mobility near to its position due to Coulomb scattering. Current fluctuations are calculated according to [3]:

$$\frac{\delta I_d}{I_d} = \frac{\delta N_c}{N_c(y)} + \frac{\delta \mu}{\mu(y)} = \left( \frac{1}{N_c(y)} \cdot \frac{\delta N_c}{\delta N_{tr}} + \frac{1}{\mu(y)} \cdot \frac{\delta \mu}{\delta N_{tr}} \right) \cdot \delta N_{tr} \qquad (4)$$

Here $N_{tr}$ is the number of traps. $\delta N_c / \delta N_{tr}$ is the change in the number of free carriers versus the number of occupied traps and $\delta \mu / \delta N_{tr}$ describes the influence of a charged trap state on the mobility at location $y$.

According to [13] the first term of eq. (4) is given by the following relation:

$$\frac{\delta N_c}{\delta N_{tr}} = \frac{q^2 / kT \cdot N_c(y)}{C_{ox} + C_{it} + C_d + q^2 / kT \cdot N_c(y)} \qquad (5)$$

$C_{it}$ is the interface trap capacitance and $C_d$ the capacitance of the pn-junction.

To derive an equation for the second term in (5), $\delta \mu / \delta N_{tr}$, we have to approximate the influence of a trap on the local mobility. The mobility is given by the inverse sum of a Coulomb-scattering related term, $\mu_C$, and an interface scattering related contribution $\mu_{SR}$ [14]:

$$\frac{1}{\mu} = \frac{1}{\mu_{SR}} + \frac{1}{\mu_C} \qquad (6)$$

According to [16], interface scattering and coulomb-scattering are approximated by

$$\mu_{SR} = \frac{\mu_{SR,0}}{\theta \left( V_{g,eff} - V(y) + Q_B / C_{ox} \right)^2} \qquad (7)$$

and

$$\mu_C = \frac{\mu_{0,C}}{\gamma \cdot N_{tr} + \beta} \qquad (8)$$

respectively, with $\mu_{SR,0}, Q_B, \theta, \beta, \gamma$, and $\mu_{0,C}$ being technology-dependent physics-based constants. Here, for simplicity we assume these parameters to be constant. Eqs. (6) - (8) finally result in:

$$\mu(y) \cong \frac{\mu_0}{1 + \theta^* \left( V_{g,eff} - V(y) \right)} \qquad (9)$$

with $\mu_0$ and $\theta^*$ being constants resulting from $\mu_{SR,0}, \theta, \gamma$, and $\beta$.

On the basis of this equation the term $\delta \mu / \delta N_t$ can be easily calculated:

$$\frac{\delta \mu}{\delta N_t} = \frac{\gamma \cdot \mu^2}{\mu_{0,C}} \cdot \frac{S}{W \cdot L} \qquad (10)$$

Here $L$ is the length of the transistor and the scattering parameter $S$ is introduces for modeling the influence on the mobility contribution of the surface roughness to the fluctuation. A detailed discussion is given in [17]. In a next step we calculate the impact of the change of the number of occupied traps on the drain current:

$$\frac{\delta I_d}{\delta N_{tr}} = \frac{V_{g,eff} \cdot g_m}{N_c(y) \cdot \mu(y)} \left( \frac{N_c(y) \cdot \mu(y)}{kT \cdot (C_{ox} + C_{it} + C_d)/q^2 + N_c(y)} + \frac{\gamma \cdot \mu(y)^2 \cdot N_c(y) \cdot S}{\mu_{0,C} \cdot W \cdot L} \right) \qquad (11)$$

Using (5), (10), and (11), we obtain for the gate voltage fluctuations:

$$\frac{\delta V_g}{\delta N_{tr}} = \frac{q}{C_{ox}} \left( \frac{V_{g,eff}}{V_1 + V_{g,eff} - V(y)} + \frac{S^* \cdot V_{g,eff}}{V_2 + V_{g,eff} - V(y)} \right) \qquad (12)$$

There, $V_1$, $V_2$ and $S^*$ are given by:

$$V_1 = \frac{kT}{q} \cdot \frac{C_{ox} + C_{it}}{C_{ox}} \qquad (13)$$

$$V_2 = \frac{1}{\theta^*} \qquad (14)$$

$$S^* = -\frac{\mu_0 \cdot \gamma}{\mu_{C,0}} \cdot \frac{C_{ox}}{q} \cdot \frac{S}{\theta^*} \qquad (15)$$

Fluctuations in the number of occupied or non-occupied trap states $\delta N_{tr}$ per transistor width and frequency are related to the Fermi-Dirac distribution $f(E)$ and the mean time constant $\tau(x,E)$ for a change in the occupation of the traps [12]:

$$\delta N_{tr}^2 = \frac{N_{rt}(x,E)}{W \cdot L} \cdot f(E) \cdot (1 - f(E)) \cdot \frac{4 \cdot \tau(x,E)}{1 + (2\pi f \tau(x,E))^2} \qquad (16)$$

Using this equation and (12), we calculate the gate voltage-related noise $dS_{Vg} (= \delta V_g^2)$ per area in the channel at location $y$ caused by traps with a distance $x$ from the interface and the energy $E$ at frequency $f$:

$$dS_{Vg}(f,x,y,E) = \frac{q^2}{C_{ox}^2 W L} \cdot \left( \frac{V_{g,eff}}{V_1 + V_{g,eff} - V(y)} + \frac{S^* \cdot V_{g,eff}}{V_2 + V_{g,eff} - V(y)} \right)^2 \cdot$$

$$N_t(E,x) \cdot f(E,x)(1 - f(E,x)) \cdot \frac{4 \cdot \tau(E,x)}{1 + (2\pi f \tau(E,x))^2} dE \, dx \, dy \qquad (17)$$

Note that this approximation is somewhat different from previous formulations of that problem [2, 3, 12, 19] since we take into account local mobility effects at different locations within the device channel. For small area devices where the integration of trap and energy densities does not describe the behavior correctly, and we have to use a discrete summation instead of the integrants $dE$, $dx$ and $dy$:

$$S(f) = \sum_{i=1}^{N_{tr}} A_i^2 \; \frac{1}{f_i} \; \frac{1}{1 + \left(\frac{f}{f_i}\right)^2} \qquad (18)$$

The parameter $f_i$ defines the corner frequency of the Lorentzian spectrum of a discrete trap with index $i$:

$$f_i = 1/2\pi\tau(E,x) \qquad (19)$$

In the following, statistical parameters for low-frequency noise behavior are derived on basis of *(18)* with $A_i$ summarizing a number of terms from (17):

$$A_i = \frac{2q^2}{\pi C_{ox}^2 W L} \left( \frac{V_{g,eff}}{V_1 + V_{g,eff} - V(y_i)} + \frac{S^* \cdot V_{g,eff}}{V_2 + V_{g,eff} - V(y_i)} \right)^2 f(E_i, x_i)(1 - f(E_i, x_i)) \cdot \Delta y_i \qquad (20)$$

Considering a larger number of small area devices, or the average behavior of smaller devices, the low frequency noise can be calculated using continuously distributed quantities like trap densities instead of discrete ones. The current set of equations derived here leads to models similar to those well know from the literature when applied to large area considerations (see appendix 1). A continuous formulation of the 1/f-noise behavior for all regions of operation as well as a reduction of the number of necessary fit parameters results from the model derived in this article.

### *3 Statistical LF-Noise Modeling*

The noise of a device itself is already a statistical parameter in time, namely the standard deviation of the drain current or, alternatively, of the equivalent gate voltage. To statistically model the variations of the noise when comparing different devices we have to identify the sources of noise voltage fluctuations. As can be seen from *(18)*, the parameters sensitive to variations are the number of traps in the active region of the device $N_{tr}$, the corner frequencies $f_i$ of the different traps, as well as the amplitude $A_i$ of the different traps. In the following, a description for the variance of each of these parameters is derived.

## 3.1 Standard deviation of the LF-noise

The number of traps $N_{tr}$ is assumed to follow a Poisson distribution. If $<N_{tr}>=N$ is the average number of traps per device in an ensemble of geometrically identical devices, the probability that $N_{Tr}$ traps are found in a particular device is given by

$$P(N_{tr}) = \frac{N^{-N_{tr}} e^{-N}}{N_{tr}!} \qquad (21)$$

In order to roughly obtain a 1/f spectrum, the time constants must be approximately uniformly distributed on a logarithmic scale [10, 11]. Since the average spectrum of large MOS devices roughly shows a 1/f-behavior it is reasonable to assume a similar distribution for the time constants. Physical processes that may lead to this distribution are e.g. discussed in [11].

The probability distribution function of the trap corner frequency $f_i$ is then given by:

$$P(f_i) = \frac{1}{\ln\left(\frac{f_{max}}{f_{min}}\right)} \cdot \frac{1}{f_i} \qquad (22)$$

The average number of traps $N$ is proportional to the active device area $W$ x $L$ and equal to

$$N = N_{dec} \ln\left(\frac{f_{max}}{f_{min}}\right) W\, L \qquad (23)$$

Here ($N_{dec}$ ln 10) is the trap density per unit area and frequency decade. The frequencies $f_{min}$ and $f_{max}$ delimit the frequency interval in which RTS is the origin of the LF-noise. $N$ is then the average number of traps with corner frequencies lying between $f_{min}$ and $f_{max}$.

In the next step, a noise model for the average noise of small area devices is developed based on statistical parameters of $A_i$.

The evaluation of the standard deviation of the average value of the noise power spectral density function $S(f)$ is:

$$\sigma_{S(f)} = \sqrt{<S_f^2> - <S_f>^2} \qquad (24)$$

The average value of the noise power spectral density $<S(f)>$ is evaluated by calculating the average value of (18) over $A_i, f_i$ and $N_{tr}$ :

$$\langle S(f) \rangle = \langle \langle \langle \sum_{i=1}^{N_{tr}} A_i^2 \frac{1}{f_i} \frac{1}{1+\left(\frac{f}{f_i}\right)^2} \rangle_{A_i} \rangle_{f_i} \rangle_{N_{tr}} \quad (25)$$

After some calculations discussed in more detail in appendix 2 the equation

$$\langle S(f) \rangle = \frac{\langle A^2 \rangle N_{dec} WL}{f} \frac{\pi}{2} \quad (26)$$

is obtained. Here $\langle A_i^2 \rangle = \langle A^2 \rangle$ is the average of the squared RTS amplitudes. This equation shows the commonly known *1/f* behavior.

Next for calculating the standard deviation, we need to calculate $\langle S(f)^2 \rangle$. We start with:

$$\langle S(f)^2 \rangle = \langle \langle \langle \sum_{i=1}^{N_{tr}} \frac{A_i^4}{f_i^2} \frac{1}{\left(1+\left(\frac{f}{f_i}\right)^2\right)^2} + \sum_{i \neq j}^{N_{tr}} \frac{A_i^2 A_j^2}{f_i f_j} \frac{1}{1+\left(\frac{f}{f_i}\right)^2} \frac{1}{1+\left(\frac{f}{f_j}\right)^2} \rangle_{A_i} \rangle_{f_i} \rangle_{N_{tr}} \quad (27)$$

After some calculations which are discussed in detail in appendix 3, we finally get the standard deviation of the noise spectral density function *S(f)* due to scattering of the parameters $A_i, f_i$, and $N_{tr}$:

$$\sigma_{S(f)}^2 = \frac{\langle A^4 \rangle N_{dec} W L}{2} \left( \frac{1}{f_{min}^2 + f^2} - \frac{1}{f_{max}^2 + f^2} \right) \quad (28)$$

Where $\langle A_i^4 \rangle = \langle A^4 \rangle$. If *f_{min}* becomes very small and *f_{max}* relatively large compared to the noise bandwidth of interest, a simplification is possible:

$$\sigma_{S(f)}^2 = \frac{\langle A^4 \rangle N_{dec} W L}{2} \frac{1}{f^2} \quad (29)$$

The normalized standard deviation amounts to:

$$\frac{\sigma_{S(f)}}{\langle S(f) \rangle} = \frac{\sqrt{2}}{\pi \sqrt{N_{dec} WL}} \sqrt{\frac{\langle A^4 \rangle}{\langle A^2 \rangle^2}} \quad (30)$$

Here, the contributions due to scattering of the parameters $A_i$, $N_{tr}$, and $f_i$ are all taken into account.

As can be seen from (29) and (30), fluctuations in the amplitude of individual RTS have a strong influence on the standard deviation of the LF-noise. Hence, the sources of fluctuations in the amplitude of individual RTS have to be investigated in more detail. This is done in the next section.

## 3.2 Statistical parameters of the LF-noise amplitude

The amplitude $A_i$ of a RTS current fluctuation results from the combined effect of carrier number and mobility fluctuation as given by (4). To model the total standard deviation of the noise power spectral density it is necessary to investigate the factors that influence $\delta N_C /\delta N_C(y)$ and $\delta\mu/\mu(y)$.

We first investigate the mobility fluctuation term $\delta\mu/\mu(y)$. The mobility is impacted by carrier scattering at the location of the traps (cf. Eq. (10)). Scattering efficiency depends on inversion layer parameters, like charge carrier velocity and carrier density, and on the device geometry. A charge closer to the interface scatters carriers more effectively than one further away [5]. If the vertical distance $d$ of the trap from the inversion layer is a random variable, it contributes to dispersion of the noise. To best of our knowledge analytical models for the scattering efficiency as a function of $d$ have not been published so far.

A reasonable first order approach is to assume the scattering efficiency to be proportional to the intersection between the channel plane and the sphere defined by the critical trap radius $r_c$, as depicted in Fig. 3. The parameter $r_c$ is assumed to be either the distance of Coulomb interaction energy which is greater than $kT$ or the screening length $L_S$. Since the Coulomb potential is

$$V(r) = q/(4\pi\varepsilon r) \quad (31)$$

with $r$ being the radial distance from the trap, the critical radius for Coulomb interaction $r_{kt}$ is given by

$$r_{kt} = \frac{q^2}{4\pi\varepsilon kT_e} \quad (32)$$

For a two-dimensional electron gas, $L_S$ is approximated by [4]:

$$L_S = \sqrt{2}\frac{\varepsilon kT}{q N_C(y)} \quad (33)$$

The critical radius $r_c$ is given by the minimum of $L_S$ and $r_{kt}$. The radius of intersection between the channel plane and the sphere defined by the critical trap radius $r_c$ is

$$r_i = \sqrt{r_c^2 - d^2} \quad (34)$$

The channel area perturbed by the trap is then given by

$$A_{trap} = \pi r_i^2 = \pi (r_c^2 - d^2) \quad (35)$$

Since $r_c$ depends on the inversion layer carrier concentration and carrier temperature, $A_{trap}$ may strongly depend on the bias point, especially at the drain side.

In a first order approach $\Delta\mu/\mu$ is estimated to be equal to the ratio between the perturbed area and the active channel area $(W x L)$:

$$\frac{\delta\mu}{\mu(y)} = \frac{A_{trap}}{W\,L} = \frac{\pi\,r_i^{\;2}}{W\,L} \quad (36)$$

It is assumed that $d$ only affects the scattering efficiency of the trap, i.e., the change in mobility $\delta\mu$. There is no correlation between $d$ and $\delta N$ as long as $d << t_{ox}$. Since this condition is true for all traps with significant contributions to the device noise, this means that we can treat mobility and number fluctuations for different devices as statistically independent parameters.

The value of $r_c$ and hence $r_i$ depend on the bias conditions in a complex manner. Consequently, it is difficult to provide a closed expression for the variance in noise power due to mobility fluctuations. Nevertheless, at small drain bias $r_c$ can be assumed to be constant at all channel positions in a first order approximation. If in addition all distances $d$ are assumed to have equal probabilities (for $0 \leq d \leq r_c$), $\sigma(\delta\mu/\mu)$ / $<\delta\mu/\mu>$ can be calculated using eq. (36):

$$\frac{\sigma(\delta\mu/\mu)}{<\delta\mu/\mu>} = \frac{1}{\sqrt{5}} = k_d \quad (37)$$

The contribution of the mobility fluctuation to the total variance in noise power due to carrier scattering is particularly important at small drain voltages. Here the channel is homogeneous and variance in $\delta N_C$ /$\delta N_C(y)$ approaches zero. Therefore, the variance due to mobility fluctuations can be modeled as a constant in a first order approximation, here.

Let us now investigate the carrier number fluctuation term $\delta N_C$ /$\delta N_C(y)$. For this purpose the carrier density $N_C(y)$ at all positions $y$ within the channel has to be known. This parameter depends on the bias condition and is a function of the local channel potential $V(y)$ within the channel, as given by (2).

Since the number of free carriers is a function of the position $y$, also the carrier number fluctuation term $\delta N_C/\delta N_C(y)$ depends on $y$. The influence of a trap on the total current through the device depends on the local number of free carriers. Therefore, also the RTS amplitude depends on the position $y$ of a trap within the channel. At low drain voltage $V_d$ there is almost no dependence on $y$, but for transistor operating points in the saturation region this effect is strongly pronounced. For small drain voltages the carrier density is approximately homogenous within the whole channel. At high $V_d$ however, the carrier density decreases from source to drain, and $\delta N_C / \delta N_C(y)$ increases from source to the pinch-off point, where it reaches its maximum. Hence, the scattering in RTS amplitudes due to the variance of $\delta N_C / \delta N_C(y)$ increases with increasing drain bias and reaches a maximum when the device is operated in saturation.

In order to evaluate $\delta N_C / \delta N_C(y)$ at different bias points a noise efficiency term $h(y)$ is introduced which describes the efficiency of a trap in producing noise related to number fluctuation [1]. The amplitude $A_i$ of the trap is then directly proportional to $h(y)$. This term depends on bias point and on the trap position, and is given by the following approximation:

$$h(y) = \frac{\partial S_{Vg}(y)}{S_{Vg}} = K_y \frac{V_{g,eff}}{kT/q \cdot (C_{ox} + C_{it})/ C_{ox} + V_{gd}(y)} \qquad (38)$$

with $h(y)$ normalized according to:

$$\int_0^L h(y) dy = 1 \qquad (39)$$

and $V_{gd}(y)$ given by:

$$V_{gd}(y) = \begin{cases} 0 & \text{for } V_{g,eff,krit} \leq V_d \text{ and } V_{g,eff,krit} > V_{g,eff} \\ V_{g,eff} - V_{g,eff,krit} & \text{for } V_{g,eff,krit} \leq V_d \text{ and } V_{g,eff,krit} < V_{g,eff} \\ V_{g,eff} - V(y) & \text{for } V_{g,eff,krit} > V_d \end{cases} \qquad (40)$$

with $V_{g,eff,krit} = v_s L / \mu$ and $v_s$ being the saturation velocity. For simplicity, the potential $V(y)$ determining the local channel carrier concentration of the MOSFET is approximated by a linear fit here:

$$V(y) = V_{g,eff} \cdot \frac{L - y}{L} + V_d \cdot \frac{y}{L} \qquad (41)$$

In eqs. (38) – (40) we also take into account the effects arising from velocity saturation in the channel. If the vertical field at a certain position in the channel is sufficiently high for free carriers to reach saturation velocity in the inversion region, their density between this point and the drain junction remains constant. Within this region, $h(y)$ remains constant. Finally, near drain and beyond the pinch-off point there is no attractive field for free carriers at the interface, so that they do not interact with traps. For this reason the impact of traps in this region on trap related noise can be neglected.

Using these approximations we can calculate the dependence of $A_i$ on bias point. To get an expression for the standard deviation we have to statistically sum up the infinitesimal small areas with different amplitudes $A_i$ to the total noise and evaluate the resulting deviation. The resulting deviation of the noise amplitude from its average value is proportional to the integral of $h(y)^2$ from source to drain. Neglecting the non-dominant terms in this integral, the normalized deviation of noise amplitude from its average amounts to [1]:

$$\frac{\sigma(\delta N_C/N_C)}{<\delta N_C/N_C>} \approx \frac{V_D^5}{V_{Geff}^5} \qquad (42)$$

This equation describes statistical noise deviations due to the bias point. It neither depends on technology parameters or on device geometry nor requires additional fit parameters. The scattering in RTS amplitudes due to the variance in bias point increases with drain bias and reach a maximum at saturation. For very small values of $V_d$ the density of carriers is homogenous along the channel, and $h(y)=1$ within the whole channel. This condition gives us minimal variation of the noise power spectral density. As $V_d$ increases the number of free carriers decreases with $y$ increasing, and $h(y)$ increases from source to drain resulting in higher variations of the noise power spectral density.

With the above assumptions the term due to fluctuations in noise amplitude can finally be written as

$$\frac{<A^4>}{<A^2>^2} = 1 + \frac{V_D^5}{V_{Geff}^5} + k_d \qquad (43)$$

In the above equation, $k_d$ describes the mobility influence and models the fluctuations in RTS amplitude that are present even at low drain bias, where $N_C(y)$ is homogenous within the whole channel and fluctuations are due to scattering of $d$. The term $V_d^5/V_{g,eff}^5$ weights the fluctuations due to the non homogeneous contribution of the traps to the low frequency noise depending on channel position at larger drain bias.

A MiniMOS [9] device simulation is performed to investigate the term $<A^4>/<A^2>^2$ under different bias conditions. In the device simulations both number and mobility fluctuations are taken into account to determine the amplitude of the drain current fluctuations $\delta I_d$. Fig. 4 shows both device simulations and the results of $h(y)$, as a function of the trap position along the channel. Good agreement between the simplified model and the result from device simulations is found.

### 3.3 Standard deviation of LF-noise for different bandwidths of interest

The noise amplitude at a given frequency $f$ and its standard deviation is an important parameter to the circuit designer. But also the noise power integrated over the circuit bandwidth, $np_{BW}$, and its related standard deviation are of interest in many cases. This parameter is given by the integration of eq. *(18)* from $f_L$ to $f_H$, the lower and upper boundaries of the bandwidth of interest in a given circuit design.

$$np_{BW} = \int_{f_L}^{f_H} S(f)\, df \qquad (44)$$

Inserting eq. *(18)* in eq. (44) leads to

$$np_{BW} = \int_{f_L}^{f_H} \left( \sum_{i=1}^{N_{tr}} A_i^2 \; \frac{1}{f_i} \; \frac{1}{1+\left(\frac{f}{f_i}\right)^2} \right) df \qquad (45)$$

After integration the average value is derived:

$$<np_{BW}> \; = \; <A^2> \frac{\pi}{2} \ln\left(\frac{f_H}{f_L}\right) N_{dec}\, WL \qquad (46)$$

Finally, the statistical variance in noise power due to scattering in $A_i$, $f_i$, and $N_{tr}$ is calculated:

$$\sigma_{np}^2 = <np_{BW}^2> - <np_{BW}>^2 = <A^4>\, N_{dec}\, WL \left( \frac{\pi}{2} \ln\left(\frac{f_H}{f_L}\right) \right)^2 \qquad (47)$$

The normalized standard deviation is then:

$$\frac{\sigma_{np}}{<np_{BW}>} = \frac{1}{\sqrt{N_{dec} WL}} \sqrt{\frac{<A^4>}{<A^2>^2}} \qquad (48)$$

The derivation of $<np_{BW}>$ and $\sigma_{np}$ is discussed in more detail in appendix 4. For the circuit design it is important to take into account that the standard deviation depends on number of traps, device geometry and bias point, and that the bandwidth dependence is weak. This is also discussed in greater detail in appendix 4.

### 3.4 Contribution of technology dependent long range statistical parameters on the standard deviation of LF-noise

Eq. (48) assumes that there is no correlation between the standard deviation and the spacing $D$ of two devices. However, experimental data reveal correlations between noise amplitude and transistor position, as shown in Fig. 7.

The long range correlation distance is considered by a parameter $S$ in the following. This parameter describes the variation of the low frequency noise as a function of the spacing $D$ [7]. Modification of eq. (48) thus leads to

$$\frac{\sigma_{np}^{2}}{<np_{BW}>^{2}} = \frac{1}{N_{dec}WL}\frac{<A^{4}>}{<A^{2}>^{2}} + S^{2}D^{2} \qquad (49)$$

The effect of $S$ is important only for large area devices at significant spacing and can be included into a compact model to simulate the long range variation effects.

## 4 Experimental

In this chapter the model proposed in the previous chapter is validated and compared to experimental data from three different CMOS technologies with minimum feature sizes of 0.25µm ($t_{ox}$=5nm, $V_{dd}$=2.5V), 0.13µm ($t_{ox}$=2.2nm, $V_{dd}$=1.5V), and 0.09µm ($t_{ox}$=1.6nm, $V_{dd}$=1.2V).

As seen in Fig. 1, the Lorentzian shape function of an individual RTS dominates the LF-noise characteristics. This effect can be used to distinguish between trap number related and amplitude related fluctuations. The number of traps $N_{tr}$ is taken from the visible Lorentz spectra in the measured frequency range. From that data $1/N_{dec}$ is calculated (Table 1). The difference between those calculations and the total fluctuations give information on other sources of fluctuations discussed in the previous chapter.

For the 0.25µm technology the number of traps $N_{tr}$ is also extracted by performing charge pumping measurements at 1 MHz. Reasonable agreement between $N_{tr}$ evaluated from LF-noise and charge-pumping data is found confirming the proposed strategy. After extraction of $N_{tr}$, the term $(<A^{4}>/<A^{2}>^{2})^{0.5}$ is calculated (Table 1). The resulting fluctuations of the normalized noise amplitude are higher than expected for the case where only number fluctuations in $N_{tr}$ are taken into account. This fact experimentally confirms that RTS-Amplitude fluctuations are relevant for the statistical fluctuations of the low frequency noise amplitude.

Figure 5 shows the normalized standard deviation of the low frequency noise of measured transistors as a function of device area. The area dependence predicted by eq. (48) is clearly observed here showing excellent agreement between experiment and model.

In long channel devices the average gate referred voltage noise is widely independent of bias conditions. However this is not true for noise fluctuations between different de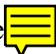ubmicron devices. As can be seen in Fig. 2, strong

fluctuations in noise performance do not only appear between different devices, but also for different bias points of a single device. Generally, fluctuations in noise amplitude increase for large gate and especially drain voltages (see Fig. 6) due to the increasing influence of the trap position in the channel. This is experimentally proven by the data shown in Fig. 6. At low $V_d$, the density of carriers is homogenous along the channel. As $V_d$ increases the number of free carriers decreases from source to drain, and the normalized standard deviation increases in the opposite direction. Again model and experiment show good agreement without any technology dependent additional fit parameter.

The distribution of average noise power spectral density across a 200 mm wafer is shown in Figure 7. This plot gives an idea about long range process related fluctuations. Analysis of long range noise amplitude fluctuation versus distance for the 0.25μm and 0.13μm nodes leads to *SD* equal to 0.18 and 0.26 (see Eq. 49), respectively.

## Conclusion

This paper discusses the impact of statistical effects on the low-frequency noise performance of CMOS devices in modern technologies. A novel low-frequency noise model including detailed physics based modeling of statistical effects is presented. Strong variations of noise performance may appear not only between devices, but also for a single device operated under different bias conditions. The noise performance is shown to depend on the number of traps, the trap position within the channel, on the depth of the trap location within the oxide, on the bias point, on device geometry, and on long-range statistical parameters. Good agreement between model and experiment is observed. The derived model can easily be implemented in a Spice like circuit simulator and is compatible with the BSIM models.

**Appendix 1: Evaluation of the average noise power spectral density using trap and energy densities**

This appendix contains a trap density based calculation of the MOSFET low frequency noise. Since trap densities imply a formulation which is based on mean values, the result of this calculation leads to an insight into the average noise and allows a more detailed description of the influence of device operating points. Model derivation is based on the approaches in chapter 2.

The integration of (17) in chapter 2 leads to:

$$S_{Vg}(f) = \frac{q^2 \, kT N_{rt}(E_f)}{4\alpha C_{ox}^2 W L^2} \cdot \frac{1}{f} \int \left( \frac{V_{g,eff}}{V_1 + V_{g,eff} - V(y)} + \frac{S^* \cdot V_{g,eff}}{V_2 + V_{g,eff} - V(y)} \right)^2 dy \qquad (50)$$

Here $\alpha$ is the tunnel parameter assuming a WKB-like tunneling behavior of the traps, and calculated similar to [12].

According to [14] the incremental location $dy$ is related to the incremental channel potential $dV$ by:

$$dy = \frac{W \cdot \mu_0 \cdot C_{ox} \cdot V_2}{g_m \cdot V_{g,eff}} \cdot \frac{V_{g,eff} - V(y)}{V_2 + V_{g,eff} - V(y)} dV \qquad (51)$$

Since the traps between drain and pinch-off point only provide a negligible contribution to the transistor's noise, the integration can be simplified by setting the pinch-off point as the upper integral boundary. Integration leads to a formulation of the low frequency noise caused by trap states at the oxide-semiconductor interface:

$$\begin{aligned}
S_{Vg}(f) = S_{Vg}^0 \cdot \frac{V_{g,eff}}{g_m} \Bigg[ & \frac{1}{V_2} \left( \ln \left( \frac{V_1 + V_{g,eff}}{V_2 + V_{g,eff}} \frac{V_2 + V_{gd}}{V_1 + V_{gd}} \right) + \frac{V_1}{V_2 + V_{g,eff}} - \frac{V_1}{V_2 + V_{gd}} \right) \\
& + 2S^* \left( \ln \left( \frac{V_1 + V_{gd}}{V_2 + V_{gd}} \frac{V_2 + V_{g,eff}}{V_1 + V_{g,eff}} \right) + \frac{1}{V_2 + V_{gd}} - \frac{1}{V_2 + V_{g,eff}} \right) \\
& + \frac{S^{*2}}{2} \left( \frac{V_2 - 2V_{gd}}{(V_2 + V_{gd})^2} - \frac{V_2 - 2V_{g,eff}}{(V_2 + V_{g,eff})^2} \right) \Bigg] \cdot \frac{1}{f}
\end{aligned} \qquad (52)$$

with

$$V_{gd} = \begin{cases} 0 & for \ V_{g,eff} \le V_d \\ V_{g,eff} - V_d & for \ V_{g,eff} > V_d \end{cases} \qquad (53)$$

and

$$S_{Vg}^0 = \frac{q^2 \, kT N_{tr}(E_f) \mu_0}{4\alpha \theta^* C_{ox} L^2} \qquad (54)$$

The model based on eq. (52) is proven to be compatible to the compact models of the BSIM standard [18]. The model formulation is similar to the BSIM sub-threshold formulation, but is continuous over the whole range of operating points of a MOSFET. For devices operating in inversion the main difference to BSIM and other analytical low frequency noise models is the consideration of local differences in the mobility at different locations within the channel. This allows to eliminate one fit parameter compared to the BSIM approach [15, 19]. The remaining free parameters are the physics related fit parameters $N_t(E_f)$ and $\mu_{C,0} / \gamma$, which describe number and mobility fluctuation related contributions to the low frequency noise. All other parameters are standard BSIM parameters.

**Appendix 2: Evaluation of the noise power spectral density $<S(f)>$ on the basis of an RTS formulation**

Here we discuss the derivation of the low frequency noise based on the formulation in chapter 2 which takes into account the discrete nature of the traps. The purpose here is to compare the approaches from appendix 1 and from this appendix. Moreover, the result of this calculation is needed for the evaluation of the higher statistical moments.

Starting from eq. (25) and observing that the amplitude $A_i$ of the RTS of the $i$th trap is for a first order approximation statistically independent of the number of traps $N_{tr}$ in a device, and also independent of the corner frequency $f_i$ [1]:

$$<S(f)> = < < \sum_{i=1}^{N_{tr}} <A_i>^2 \frac{1}{f_i} \frac{1}{1+\left(\frac{f}{f_i}\right)^2} >_{f_i} >_{N_{tr}} \qquad (55)$$

In the following $<A_i^2>$ is written as $<A^2>$. To evaluate the average over $f_i$, we use (22) for $P(f_i)$ leading to

$$<S(f)> = \sum_{i=1}^{N_{tr}} \frac{<A^2>}{\ln\left(\frac{f_{max}}{f_{min}}\right)} \int_{fmin}^{fmax} \frac{1}{f_i^2} \frac{df_i}{1+\left(\frac{f}{f_i}\right)^2} \qquad (56)$$

After some mathematics we obtain:

$$<S(f)> = \sum_{i=1}^{N_{tr}} \frac{<A^2>}{\ln\left(\frac{f_{max}}{f_{min}}\right)} \frac{1}{f}\left( \arctan\left(\frac{f_{max}}{f}\right) - \arctan\left(\frac{f_{min}}{f}\right) \right) \qquad (57)$$

Finally, we have to consider the average of $N_{tr}$:

---

[1] The amplitude and also the corner frequency depend on the trap position in the channel. This is discussed in more detail in section 3.2. However the correlation between these parameters is weak and can therefore be neglected here for simplicity.

$$\langle S(f)\rangle = \langle \sum_{i=1}^{N_{tr}} \frac{\langle A^2\rangle}{\ln\left(\frac{f_{max}}{f_{min}}\right)} \frac{1}{f}\left(\arctan\left(\frac{f_{max}}{f}\right) - \arctan\left(\frac{f_{min}}{f}\right)\right)\rangle_{N_{tr}} \quad (58)$$

$$\langle S(f)\rangle = \sum_{N_{tr}=0}^{\infty} \left(\sum_{i=1}^{N_{tr}} \frac{\langle A^2\rangle}{\ln\left(\frac{f_{max}}{f_{min}}\right)} \frac{1}{f}\left(\arctan\left(\frac{f_{max}}{f}\right) - \arctan\left(\frac{f_{min}}{f}\right)\right)\right) P(N_{tr}) \quad (59)$$

The summation from $i=1$ to $i=N_{tr}$ leads to

$$\langle S(f)\rangle = \sum_{N_{tr}=0}^{\infty} \left(\frac{\langle A^2\rangle}{\ln\left(\frac{f_{max}}{f_{min}}\right)} \frac{1}{f}\left(\arctan\left(\frac{f_{max}}{f}\right) - \arctan\left(\frac{f_{min}}{f}\right)\right) N_{tr}\right) P(N_{tr}) \quad (60)$$

Inserting eq. (21) for $P(Ntr)$ leads to

$$\langle S(f)\rangle = \sum_{N_{tr}=0}^{\infty} \left(\frac{\langle A^2\rangle}{\ln\left(\frac{f_{max}}{f_{min}}\right)} \frac{1}{f}\left(\arctan\left(\frac{f_{max}}{f}\right) - \arctan\left(\frac{f_{min}}{f}\right)\right) N_{tr}\right) \frac{N^{-N_{tr}} e^{-N}}{N_{tr}!} \quad (61)$$

$$\langle S(f)\rangle = \left(\frac{\langle A^2\rangle}{\ln\left(\frac{f_{max}}{f_{min}}\right)} \frac{1}{f}\left(\arctan\left(\frac{f_{max}}{f}\right) - \arctan\left(\frac{f_{min}}{f}\right)\right)\right) \sum_{N_{tr}=0}^{\infty} N_{tr} \frac{N^{-N_{tr}} e^{-N}}{N_{tr}!} \quad (62)$$

and finally to:

$$\langle S(f)\rangle = \left(\frac{\langle A^2\rangle}{\ln\left(\frac{f_{max}}{f_{min}}\right)} \frac{1}{f}\left(\arctan\left(\frac{f_{max}}{f}\right) - \arctan\left(\frac{f_{min}}{f}\right)\right)\right) \langle N_{tr}\rangle \quad (63)$$

$\langle S(f)\rangle$ together with eq. (23) can be rewritten as:

$$\langle S(f)\rangle = \frac{\langle A^2\rangle N_{dec} WL}{f}\left(\arctan\left(\frac{f_{max}}{f}\right) - \arctan\left(\frac{f_{min}}{f}\right)\right) \quad (64)$$

If $f_{min}$ is much smaller than the lower frequency at which the 1/f noise is of practical interest, and if $f_{max}$ is much higher than the frequency at which the thermal noise supersedes 1/f noise, this results in

$$\langle S(f)\rangle = \frac{\langle A^2\rangle N_{dec} WL}{f} \frac{\pi}{2} \quad (65)$$

Note that a proportionality to $1/WL$ and to the average number of traps is obtained here similar to eq. (52). This is because the number of traps $N_{tr}$ in the device is equal to $N_{dec}$ x $W$ x L and because $A$ is proportional to $1/WL$. The equation above can be rewritten:

$$<S(f)> = \frac{k\ f_1(Vd,\ Vg)\ N}{WL\ f} \quad (66)$$

Here $f_1(V_d, V_g)$ contains the bias point dependence hidden in the parameter $<A^2>$ in the above equation and $k$ is a constant. Therefore the final result is equivalent to eq. (52) and therefore also to the standard low frequency noise models used for BSIM formulation [19]. The main difference is the use of a microscopic formulation of the low frequency noise which helps determining its statistical behavior. As already mentioned this derivation in contrast to appendix 1 does not detail the dependence of the low frequency noise on the operating point[2], but shows that the statistical approach used in chapter 3 is equivalent to the results for the average noise of large area devices. Note that this is not a shortcoming of this formulation, but it is a problem which is of minor importance here, since the fluctuations in the noise are much higher than the dependence of the average value on bias point. Moreover, the dependence of the fluctuations in amplitude on bias point is properly modeled by eq. (43).

### Appendix 3: Evaluation of the standard deviation of the noise power spectral density $<S(f)>$

In this appendix the detailed calculation of the standard deviation of the noise power spectral density of MOSFETs is discussed.

$<S(f)>$ has been calculated in appendix 2. In the following $<S(f)^2>$ is calculated. Starting from eq. (27) and assuming that the distribution of amplitudes $A_i$ is statistically independent of the distribution of $N_{tr}$ and $f_i$ (see footnote in appendix 2) and finally defining some average values: $<A_i^4> = <A^4>$ and $<A_i^2\ A_j^2> = <A^2>^2$, we may rewrite this equation as:

$$<S(f)^2> = << \sum_{i=1}^{N_{tr}} \frac{<A^4>}{f_i^2} \frac{1}{\left(1+\left(\frac{f}{f_i}\right)^2\right)^2} >_{f_i} >_{N_{tr}} + << \sum_{i \neq j}^{N_{tr}} \frac{<A^2>^2}{f_i\ f_j} \frac{1}{1+\left(\frac{f}{f_i}\right)^2} \frac{1}{1+\left(\frac{f}{f_j}\right)^2} >_{f_i} >_{N_{tr}} \quad (67)$$

Thus, we have to find the average of two summands in eq. (67). First, we consider the first term on the right hand side:

---

[2] This dependence is hidden in the parameter $<A^2>$ in eq. (65)

$$\left\langle \frac{\langle A^4 \rangle}{f_i^2} \frac{1}{\left(1+\left(\frac{f}{f_i}\right)^2\right)^2} \right\rangle_{fi} = \int_{f_{min}}^{f_{max}} \frac{\langle A^4 \rangle \, P(f_i) \, df_i}{f_i^2 \left(1+\left(\frac{f}{f_i}\right)^2\right)^2} = \frac{\langle A^4 \rangle}{\ln\left(\frac{f_{max}}{f_{min}}\right)} \int_{f_{min}}^{f_{max}} \frac{df_i}{f_i^3 \left(1+\left(\frac{f}{f_i}\right)^2\right)^2}$$

$$= \frac{\langle A^4 \rangle}{\ln\left(\frac{f_{max}}{f_{min}}\right)} \frac{1}{2} \left( \frac{1}{f_{min}^2 + f^2} - \frac{1}{f_{max}^2 + f^2} \right) \quad (68)$$

Next, we evaluate the second term on the right hand side:

$$\left\langle \frac{\langle A^2 \rangle}{f_i} \frac{1}{1+\left(\frac{f}{f_i}\right)^2} \frac{\langle A^2 \rangle}{f_j} \frac{1}{1+\left(\frac{f}{f_j}\right)^2} \right\rangle_{fi} = \left\langle \frac{\langle A^2 \rangle}{f_i} \frac{1}{1+\left(\frac{f}{f_i}\right)^2} \right\rangle \left\langle \frac{\langle A^2 \rangle}{f_j} \frac{1}{1+\left(\frac{f}{f_j}\right)^2} \right\rangle$$

$$= \frac{\langle A^2 \rangle^2 \, N_{dec}^2}{f^2} \left( \arctan\left(\frac{f_{max}}{f}\right) - \arctan\left(\frac{f_{min}}{f}\right) \right)^2 \quad (69)$$

Finally, this results in:

$$\langle S(f)^2 \rangle = \left\langle \sum_{i=1}^{N_{tr}} \frac{\langle A^4 \rangle}{\ln\left(\frac{f_{max}}{f_{min}}\right)} \frac{1}{2} \left( \frac{1}{f_{min}^2 + f^2} - \frac{1}{f_{max}^2 + f^2} \right) + \right.$$

$$\left. + \sum_{i \neq j}^{N_{tr}} \frac{\langle A^2 \rangle^2 \, N_{dec}^2}{f^2} \left( \arctan\left(\frac{f_{max}}{f}\right) - \arctan\left(\frac{f_{min}}{f}\right) \right)^2 \right\rangle_{N_{tr}} \quad (70)$$

This sum is rewritten as follows:

$$\langle S(f)^2 \rangle = \left\langle \frac{\langle A^4 \rangle}{\ln\left(\frac{f_{max}}{f_{min}}\right)} \frac{1}{2} \left( \frac{1}{f_{min}^2 + f^2} - \frac{1}{f_{max}^2 + f^2} \right) N_{tr} + \right.$$

$$\left. + \frac{\langle A^2 \rangle^2 \, N_{dec}^2}{f^2} \left( \arctan\left(\frac{f_{max}}{f}\right) - \arctan\left(\frac{f_{min}}{f}\right) \right)^2 N_{tr}\,(N_{tr}-1) \right\rangle_{N_{tr}} \quad (71)$$

Since $N_{tr}$ follows a Poisson distribution the average is:

$$\langle S(f)^2 \rangle = \frac{\langle A^4 \rangle}{\ln\left(\frac{f_{max}}{f_{min}}\right)} \frac{1}{2} \left( \frac{1}{f_{min}^2 + f^2} - \frac{1}{f_{max}^2 + f^2} \right) \sum_{N_{tr}=0}^{\infty} N_{tr} \frac{N^{-N_{tr}} \, e^{-N}}{N_{tr}!} +$$

$$+ \frac{<A^2>^2 N_{dec}^2}{f^2}\left(\arctan\left(\frac{f_{max}}{f}\right) - \arctan\left(\frac{f_{min}}{f}\right)\right)^2 \sum_{N_{tr}=0}^{\infty} N_{tr}(N_{tr}-1)\frac{N^{-N_{tr}}e^{-N}}{N_{tr}!} \quad (72)$$

For a Poisson distribution the following relations hold:

$$\sum_{N_{tr}=0}^{\infty} N_{tr}\frac{N^{N_{tr}}e^{-N}}{N_{tr}!} = N \quad \text{and} \quad \sum_{N_{tr}=0}^{\infty} N_{tr}^2 \frac{N^{N_{tr}}e^{-N}}{N_{tr}!} = N(N+1)$$

Therefore we finally achieve:

$$<S(f)^2> = \frac{<A^4> N_{dec}\, W\, L}{\ln\left(\frac{f_{max}}{f_{min}}\right)}\frac{1}{2}\left(\frac{1}{f_{min}^2+f^2} - \frac{1}{f_{max}^2+f^2}\right) \; +$$

$$+ \frac{<A^2>^2 N_{dec}^2 (WL)^2}{f^2}\left(\arctan\left(\frac{f_{max}}{f}\right) - \arctan\left(\frac{f_{min}}{f}\right)\right)^2 \quad (73)$$

From this and appendix 2 we calculate the standard deviation:

$$\sigma_{S(f)}^2 = \frac{<A^4> N_{dec}\, W\, L}{2}\left(\frac{1}{f_{min}^2+f^2} \; - \; \frac{1}{f_{max}^2+f^2}\right) \quad (74)$$

If $f_{min}$ is much smaller than the lower frequency at which the 1/f noise is of practical interest, and if $f_{max}$ is much higher than the frequency at which the thermal noise supersedes 1/f noise, this results in:

$$\sigma_{S(f)}^2 = \frac{<A^4> N_{dec}\, W\, L}{2}\frac{1}{f^2} \quad (75)$$

Finally, the normalized standard deviation amounts to:

$$\frac{\sigma_{S(f)}}{<S(f)>} = \frac{\sqrt{2}}{\pi\sqrt{N_{dec} WL}}\sqrt{\frac{<A^4>}{<A^2>^2}} \quad (76)$$

### Appendix 4: Evaluation of the noise power in the bandwidth of interest and its standard deviation

In this appendix a statistical analysis of the standard deviation of $np_{BW}$ depending on $f_L$ and $f_H$ is given. Starting point will be eq. (45). The goal is to provide a compact model as a function of a minimum set of geometrical and technological parameters.

The integral in eq. (45) can be solved to:

$$np_{BW} = \sum_{i=1}^{N_{tr}} A_i^2 \text{ arc tg} \left(\frac{f_H}{f_i}\right) - \sum_{i=1}^{N_{tr}} A_i^2 \text{arc tg} \left(\frac{f_L}{f_i}\right) = \sum_{i=1}^{N_{tr}} np_i \qquad (77)$$

Here $np_i$ is the contribution of a single trap with corner frequency $f_i$ and amplitude $A_i$ to the noise power integrated over the bandwidth. The total noise power is the sum of the contribution of all traps. Notice that even if the corner frequency $f_i$ lies outside the bandwidth delimited by $f_L$ and $f_H$ it does contribute to the noise power in the bandwidth, according to the equation above.

In the following we will evaluate both average value and standard deviation of a larger ensemble of nominally identically transistors (but with different statistically distributed traps). We start with the calculation of the average based on eq. (77).

If $np_{BW}(N_{tr})$ is the noise power for the number of traps in the device to be equal to $N_{tr}$, and $P(N_{tr})$ is the probability that the number of traps in the device is equal to $N_{tr}$, then

$$<np_{BW}> = \sum_{N_{tr}=0}^{\infty} np_{BW}(N_{tr}) \, P(N_{tr}) \quad (78)$$

Here $np_{BW}(N_{tr})$ is given by eq. (45) and $P(N_{tr})$ follows a Poisson distribution. Hence,

$$<np_{BW}> = \sum_{N_{tr}=0}^{\infty} \sum_{i=1}^{N_{tr}} np_i \frac{N^{-N_{tr}} e^{-N}}{N_{tr}!} = \sum_{N_{tr}=0}^{\infty} (<np_i> N_{tr}) \frac{N^{-N_{tr}} e^{-N}}{N_{tr}!} \quad (79)$$

Here $N=<N_{tr}>$ is the average number of traps.

Let us first investigate the average of $np_i$ given by:

$$<np_i> = \int_{f_{min}}^{f_{max}} np_i(A_i, f_i).P(f_i).df_i = \frac{1}{\ln\left(\frac{f_{max}}{f_{min}}\right)} \int_{f_{min}}^{f_{max}} \left(A_i^2 \text{ arc tg}\left(\frac{f_H}{f_i}\right) - A_i^2 \text{arc tg}\left(\frac{f_L}{f_i}\right)\right) \frac{1}{f_i} \, df_i \quad (80)$$

Here $f_{min}$ and $f_{max}$ delimit the frequency interval in which RTS is the origin of the low-frequency noise. Note that those frequencies are different from $f_L$ and $f_H$, which are the boundaries of the bandwidth of interest in a given circuit design.

Assuming $<A_i^2> = <A^2>$,

$$<np_i> = \frac{<A^2>}{\ln\left(\frac{f_{max}}{f_{min}}\right)}\left(\int_{f_{min}}^{f_{max}} arc\,tg\left(\frac{f_H}{f_i}\right)\frac{1}{f_i}\,df_i - \int_{f_{min}}^{f_{max}} arc\,tg\left(\frac{f_L}{f_i}\right)\frac{1}{f_i}\,df_i\right) = \frac{<A^2>}{\ln\left(\frac{f_{max}}{f_{min}}\right)}(v(f_H)-v(f_L)) \quad (81)$$

The parameter $v(f_H) - v(f_L)$ is evaluated in the following:

With the definition of $u=f_H/f_i \rightarrow du = (-f_H/f_i^2)df_i \rightarrow df/f_i = (-f_i/f_H)du = (-1/u)du$

$$v(f_H) - v(f_L) = \int_{\frac{f_H}{f_{max}}}^{\frac{f_H}{f_{min}}} \frac{\arctan u}{u}\,du - \int_{\frac{f_L}{f_{max}}}^{\frac{f_L}{f_{min}}} \frac{\arctan u}{u}\,du \quad (82)$$

In this form the integral has no analytical solution, but some manipulations can be done.

Since we will always have $f_L/f_{max} < f_H/f_{max} < f_L/f_{min} < f_H/f_{min}$, $v(f_H) - v(f_L)$ becomes:

$$v(f_H) - v(f_L) = \int_{\frac{f_L}{f_{min}}}^{\frac{f_H}{f_{min}}} \frac{\arctan u}{u}\,du - \int_{\frac{f_L}{f_{max}}}^{\frac{f_H}{f_{max}}} \frac{\arctan u}{u}\,du \quad (83)$$

Since $(f_L/f_{max}) < 1$ and $(f_H/f_{max}) < 1$, as well as $(f_L/f_{min}) > 1$ and $(f_H/f_{min}) > 1$, $arctan(u)$ can now be expanded through its Taylor series leading to

$$v(f_H) - v(f_L) = \int_{\frac{f_L}{f_{min}}}^{\frac{f_H}{f_{min}}}\left(\frac{\pi}{2} + \sum_{n=0}^{\infty}\frac{(-1)^{n+1}}{(2n+1)u^{2n+1}}\right)\frac{du}{u} - \int_{\frac{f_L}{f_{max}}}^{\frac{f_H}{f_{max}}}\sum_{n=0}^{\infty}\frac{(-1)^n u^{2n+1}}{(2n+1)}\frac{du}{u} \quad (84)$$

$$v(f_H) - v(f_L) = \frac{\pi}{2}\ln\left(\frac{f_H}{f_L}\right) + \sum_{n=0}^{\infty}\frac{(-1)^{n+1}}{2n+1}\int_{\frac{f_L}{f_{min}}}^{\frac{f_H}{f_{min}}}\frac{du}{u^{2n+2}} - \sum_{n=0}^{\infty}\frac{(-1)^n}{2n+1}\int_{\frac{f_L}{f_{max}}}^{\frac{f_H}{f_{max}}} u^{2n}\,du \quad (85)$$

$$v(f_H) - v(f_L) = \frac{\pi}{2}\ln\left(\frac{f_H}{f_L}\right) + \sum_{n=0}^{\infty}\frac{(-1)^n}{(2n+1)^2}\frac{1}{u^{2n+1}}\Bigg|_{\frac{f_L}{f_{min}}}^{\frac{f_H}{f_{min}}} - \sum_{n=0}^{\infty}\frac{(-1)^n}{2n+1}\frac{u^{2n+1}}{2n+1}\Bigg|_{\frac{f_L}{f_{max}}}^{\frac{f_H}{f_{max}}} \quad (86)$$

$$v(f_H) - v(f_L) = \frac{\pi}{2}\ln\left(\frac{f_H}{f_L}\right) + \sum_{n=0}^{\infty}\frac{(-1)^n}{(2n+1)^2}\left(\left(\frac{f_{min}}{f_H}\right)^{2n+1} - \left(\frac{f_{min}}{f_L}\right)^{2n+1} - \left(\frac{f_H}{f_{max}}\right)^{2n+1} + \left(\frac{f_L}{f_{max}}\right)^{2n+1}\right) \quad (87)$$

If $f_{min} << f_L$ and $f_{max} >> f_H$, which is usually the case, the above equation can be approximated to:

$$v(f_H) - v(f_L) = \frac{\pi}{2} \ln\left(\frac{f_H}{f_L}\right) \quad (88)$$

This result gives us a simpler description of eq. (80):

$$\langle np_i \rangle = \frac{\langle A^2 \rangle \, (\, v(f_H) - v(f_L) \,)}{\ln\left(\frac{f_{max}}{f_{min}}\right)} = \frac{A^2 \, (\pi/2) \, \ln(f_H/f_L)}{\ln\left(\frac{f_{max}}{f_{min}}\right)} \quad (89)$$

Using eq. (89) in eq. (79) we get:

$$\langle np_{BW} \rangle = \sum_{N_{tr}=0}^{\infty} (N_{tr} \langle np_i \rangle) \frac{N^{-N_{tr}} e^{-N}}{N_{tr}!} = \sum_{N_{tr}=0}^{\infty} (N_{tr} \frac{A^2 \, (\pi/2) \, \ln(f_H/f_L)}{\ln\left(\frac{f_{max}}{f_{min}}\right)}) \frac{N^{-N_{tr}} e^{-N}}{N_{tr}!} \quad (90)$$

$$\langle np_{BW} \rangle = \frac{A^2 \, (\pi/2) \, \ln(f_H/f_L)}{\ln\left(\frac{f_{max}}{f_{min}}\right)} \sum_{N_{tr}=0}^{\infty} N_{tr} \frac{N^{-N_{tr}} e^{-N}}{N_{tr}!} = \frac{A^2 \, (\pi/2) \, \ln(f_H/f_L)}{\ln\left(\frac{f_{max}}{f_{min}}\right)} N \quad (91)$$

Using eq. (23) in eq. (91) we finally get

$$\langle np_{BW} \rangle = \langle A^2 \rangle \frac{\pi}{2} \ln\left(\frac{f_H}{f_L}\right) N_{dec} \, WL \quad (92)$$

for the average noise power spectral density in the frequency band between $f_L$ and $f_H$.

The calculation of the standard deviation of the noise power in the bandwidth of interest $\sigma_{np}$, is given by the following formula:

$$\sigma_{np} = \sqrt{\langle np_{BW}^2 \rangle - \langle np_{BW} \rangle^2} \quad (93)$$

$\langle np_{BW} \rangle$ is already evaluated above. So we need to evaluate $\langle np_{BW}^2 \rangle$ starting from:

$$\langle np_{BW}^2 \rangle = \left\langle \left\langle \left\langle \left( \int_{f_L}^{f_H} S(f) \, df \right)^2 \right\rangle_A \right\rangle_f \right\rangle_{N_{tr}} \quad (94)$$

$$\langle np_{BW}^2 \rangle = \left\langle \left\langle \left\langle \int_{f_L}^{f_H} \int_{f_L}^{f_H} S(f) \, S(f`) \, df \, df` \right\rangle_A \right\rangle_f \right\rangle_{N_{tr}} \quad (95)$$

$$\langle np_{BW}^2 \rangle = \int_{f_L}^{f_H} \int_{f_L}^{f_H} \left\langle \left\langle \left\langle S(f) \, S(f`) \right\rangle_A \right\rangle_f \right\rangle_{N_{tr}} df \, df` \quad (96)$$

Before the integral can be calculated, it is necessary to evaluate $<<<S(f)S(f')>>>$:

$$<S(f)\,S(f')> = <<<\sum_{i=1}^{N_{tr}}\sum_{j=1}^{N_{tr}}\frac{A_i^2}{f_i}\frac{A_j^2}{f_j}\frac{1}{\left(1+\left(\frac{f}{f_i}\right)^2\right)\left(1+\left(\frac{f'}{f_j}\right)^2\right)}>_A>_{f'}>_{N_{tr}} \quad (97)$$

$$=<\sum_{i=1}^{N_{tr}}\frac{<A_i^4>}{\ln\left(\frac{f_{max}}{f_{min}}\right)}\int_{f_{min}}^{f_{max}}\frac{f_i\,df_i}{(f_i^2+f^2)\,(f_i^2+f'^2)}+\sum_{i\neq j=1}^{N_{tr}}\frac{<A_i^2>}{\ln\left(\frac{f_{max}}{f_{min}}\right)}\int_{f_{min}}^{f_{max}}\frac{1}{f_i^2}\frac{df_i}{1+\left(\frac{f}{f_i}\right)^2}\frac{<A_j^2>}{\ln\left(\frac{f_{max}}{f_{min}}\right)}\int_{f_{min}}^{f_{max}}\frac{1}{f_j^2}\frac{df_j}{1+\left(\frac{f'}{f_j}\right)^2}>_{N_{tr}}$$

$$=\frac{<N_{tr}><A^4>}{\ln\left(\frac{f_{max}}{f_{min}}\right)(f'^2-f^2)}\ln\left(\frac{f'}{f}\right)+\left(\frac{<A^2>}{\ln\left(\frac{f_{max}}{f_{min}}\right)}\right)^2\frac{<N_{tr}(N_{tr}-1)>}{f\,f'}\left(\arctan\left(\frac{f_{max}}{f}\right)-\arctan\left(\frac{f_{min}}{f}\right)\right)\left(\arctan\left(\frac{f_{max}}{f'}\right)-\arctan\left(\frac{f_{min}}{f'}\right)\right) \quad (98)$$

$$<np_{BW}^2> = \frac{<N_{tr}><A^4>}{\ln\left(\frac{f_{max}}{f_{min}}\right)}\int_{f_L}^{f_H}\int_{f_L}^{f_H}\frac{1}{(f'^2-f^2)}\ln\left(\frac{f'}{f}\right)df\,df' +$$

$$+\int_{f_L}^{f_H}\int_{f_L}^{f_H}\left(\frac{<A^2>}{\ln\left(\frac{f_{max}}{f_{min}}\right)}\right)^2\frac{<N_{tr}(N_{tr}-1)>}{f\,f'}\left(\arctan\left(\frac{f_{max}}{f}\right)-\arctan\left(\frac{f_{min}}{f}\right)\right)\left(\arctan\left(\frac{f_{max}}{f'}\right)-\arctan\left(\frac{f_{min}}{f'}\right)\right)df\,df' \quad (99)$$

The first term on the right hand side in the above equation has no known analytical solution. Evaluation of the integral for the second term leads to

$$<np_{BW}^2> = \frac{<N_{tr}><A^4>}{\ln\left(\frac{f_{max}}{f_{min}}\right)}\int_{f_L}^{f_H}\int_{f_L}^{f_H}\frac{1}{(f'^2-f^2)}\ln\left(\frac{f'}{f}\right)df\,df' + <N_{tr}(N_{tr}-1)><A^2>^2\frac{\pi^2}{4}\ln^2\frac{f_H}{f_L} \quad (100)$$

For a Poisson distribution $<Ntr\,(Ntr-1)> = <Ntr>^2$. Then

$$<np_{BW}^2> = \frac{<N_{tr}><A^4>}{\ln\left(\frac{f_{max}}{f_{min}}\right)}\int_{f_L}^{f_H}\int_{f_L}^{f_H}\frac{1}{(f'^2-f^2)}\ln\left(\frac{f'}{f}\right)df\,df' + <N_{tr}>^2<A^2>^2\frac{\pi^2}{4}\ln^2\frac{f_H}{f_L} \quad (101)$$

This gives us the standard deviation of the noise power spectral density in the frequency band between $f_L$ and $f_H$.

$$\sigma_{np} = \sqrt{<np_{BW}^2>-<np_{BW}>^2} = \sqrt{\frac{<N_{tr}><A^4>}{\ln\left(\frac{f_{max}}{f_{min}}\right)}\int_{f_L}^{f_H}\int_{f_L}^{f_H}\frac{1}{(f'^2-f^2)}\ln\left(\frac{f'}{f}\right)df\,df'} \quad (102)$$

The normalized standard deviation is:

$$\frac{\sigma_{np}}{<np_{BW}>} = \frac{1}{\sqrt{N_{dec}WL}} \sqrt{\frac{<A^4>}{<A^2>^2}} \sqrt{\frac{\int_{f_L}^{f_H} \int_{f_L}^{f_H} \frac{1}{(f^2 - f'^2)} \ln(\frac{f'}{f}) \, df \, df'}{\ln^2 \frac{f_H}{f_L}}} \quad (103)$$

For circuit simulation purposes, further simplification is mandatory. If we simplify $<np_{BW}^2>$:

$$<np_{BW}^2> = \int_{f_L}^{f_H} \int_{f_L}^{f_H} <S(f)> <S(f')> \, df \, df' \quad (104)$$

$$<np_{BW}^2> = \int_{f_L}^{f_H} \int_{f_L}^{f_H} < \sum_{i=1}^{N_{tr}} \frac{A_i^2}{f_i} \frac{1}{1+\left(\frac{f}{f_i}\right)^2} > < \sum_{j=1}^{N_{tr}} \frac{A_j^2}{f_j} \frac{1}{1+\left(\frac{f'}{f_j}\right)^2} > \, df \, df' \quad (105)$$

The evaluation of the average over $f_i$ leads to

$$<\sum_{i=1}^{N_{tr}} \frac{A_i^2}{f_i} \frac{1}{1+\left(\frac{f}{f_i}\right)^2} >_{fi} = \sum_{i=1}^{N_{tr}} A_i^2 \int_{f_{min}}^{f_{max}} \frac{1}{f_i} \frac{P(f_i) \, df_i}{1+\left(\frac{f}{f_i}\right)^2} = \sum_{i=1}^{N_{tr}} \frac{A_i^2}{\ln(\frac{f_{max}}{f_{min}})} \int_{f_{min}}^{f_{max}} \frac{1}{f_i^2} \frac{df_i}{1+\left(\frac{f}{f_i}\right)^2} \quad (106)$$

$$= \sum_{i=1}^{N_{tr}} \frac{A_i^2}{\ln(\frac{f_{max}}{f_{min}})} \frac{1}{f^2} \int_{f_{min}}^{f_{max}} \frac{df_i}{1+\left(\frac{f_i}{f}\right)^2} = \sum_{i=1}^{N_{tr}} \frac{A_i^2}{\ln(\frac{f_{max}}{f_{min}})} \frac{1}{f} \left( \arctan(\frac{f_{max}}{f}) - \arctan(\frac{f_{min}}{f}) \right) \quad (107)$$

After the evaluation of the average over $f_j$, (105) is written as

$$<np_{BW}^2> = \int_{f_L}^{f_H} \int_{f_L}^{f_H} < \left( \sum_{i=1}^{N_{tr}} \frac{A_i^2}{\ln(\frac{f_{max}}{f_{min}})} \frac{1}{f} \left( \arctan(\frac{f_{max}}{f}) - \arctan(\frac{f_{min}}{f}) \right) \right) \cdot$$

$$\cdot \left( \sum_{j=1}^{N_{tr}} \frac{A_j^2}{\ln(\frac{f_{max}}{f_{min}})} \frac{1}{f'} \left( \arctan(\frac{f_{max}}{f'}) - \arctan(\frac{f_{min}}{f'}) \right) \right) > \, df \, df' \quad (108)$$

Observing that $<A_i^4> = <A^4>$ and $<A_i^2 A_j^2> = <A^2>^2$ the average over $A$ leads to

$$<np_{BW}^2> = \int_{f_L}^{f_H} \int_{f_L}^{f_H} < \left( N_{tr}<A^4> + N_{tr}(N_{tr}-1)<A^2>^2 \right) \left( \frac{1}{\ln(\frac{f_{max}}{f_{min}})} \frac{1}{f} \left( \arctan(\frac{f_{max}}{f}) - \arctan(\frac{f_{min}}{f}) \right) \right) \cdot$$

$$\cdot \left( \frac{1}{\ln \left( \frac{f_{max}}{f_{min}} \right)} \frac{1}{f'} \left( \arctan \left( \frac{f_{max}}{f'} \right) - \arctan \left( \frac{f_{min}}{f'} \right) \right) \right) \!\!\! >_{Ntr} df\, df' \quad (109)$$

Noting that $<N_{tr}>^2 = <N_{tr}{}^2> - <N_{tr}>$ this leads to

$$<np_{BW}{}^2> = \int_{f_L}^{f_H} \int_{f_L}^{f_H} \left( <N_{tr}><A^4> + <N_{tr}>^2 <A^2>^2 \right) \left( \frac{1}{\ln \left( \frac{f_{max}}{f_{min}} \right)} \frac{1}{f} \left( \arctan \left( \frac{f_{max}}{f} \right) - \arctan \left( \frac{f_{min}}{f} \right) \right) \right) \cdot$$

$$\cdot \left( \frac{1}{\ln \left( \frac{f_{max}}{f_{min}} \right)} \frac{1}{f'} \left( \arctan \left( \frac{f_{max}}{f'} \right) - \arctan \left( \frac{f_{min}}{f'} \right) \right) \right) df\, df' \quad (110)$$

Now the double integral from $f_L$ to $f_H$ can be evaluated in the same way as in appendix 2, leading to

$$<np_{BW}{}^2> = \left( N_{tr}<A^4> + <N_{tr}>^2 <A^2>^2 \right) \left( \frac{\pi}{2} \right)^2 \left( \frac{1}{\ln \left( \frac{f_{max}}{f_{min}} \right)} \right)^2 \left( \ln \left( \frac{f_H}{f_L} \right) \right)^2 \quad (111)$$

Finally the standard deviation is calculated:

$$\sigma_{np} = \sqrt{<np_{BW}{}^2> - <np_{BW}>^2} = \sqrt{<A^4>} \, \frac{\pi}{2} \ln \left( \frac{f_H}{f_L} \right) \sqrt{N_{dec} \, WL} \quad (112)$$

The normalized standard deviation is:

$$\frac{\sigma_{np}}{<np_{BW}>} = \frac{1}{\sqrt{N_{dec} WL}} \sqrt{\frac{<A^4>}{<A^2>^2}} \quad (113)$$

This simplified equation for the normalized standard deviation of the noise power in the bandwidth of interest shows the same dependency on geometry, average number of traps and RTS amplitude as the exact equation (103). The dependence on bias point is also the same in both equations. The major difference is the term including the integral in the second square root which cannot be analytically solved. Numerical analysis was performed and this term shows a weak dependency on $f_L$ and $f_H$. The standard deviation slightly decreases as circuit bandwidth increases. However, the numerical analysis shows that (113) is a good approximation for (103). Eq. (113) slightly overestimates the exact value given by (103). Nevertheless, eq. (113) is appropriate for circuit simulation purposes, since it correctly describes the dependency on geometry, average number of traps and RTS amplitude.

**References:**

[1] R. Brederlow, W Weber, D Schmitt-Landsidel and R Thewes, "Fluctuations of the Low Frequency Noise of MOS Transistors and their Modeling in Analog and RF-Circuits", in *IEDM Tech. Dig.*, 1999, pp.159-162.

[2] T. Boutchacha and G. Ghibaudo, "Low Frequency Noise Characterization of 0.18μm Si CMOS Transistors", *Phys. Stat. Sol. (a)* , vol. 167, pp. 261-270, May 1998.

[3] R. Jayaraman, and C.G. Sodini, "A 1/f Noise Technique to Extract the Oxide Trap Density Near the Conduction Band Edge of Silicon", *IEEE Trans. El. Dev.*, vol. 36, pp. 1773-1782, Sept. 1989.

[4] E. Simoen, B Dierickx, C L Clayes and G J Declerck, "Explaining the Amplitude of RTS Noise in Submicrometer MOSFETs", *IEEE Trans. El. Dev.*, vol. 39, pp.422-419, Feb. 1992.

[5] A. Godoy, F Gámiz, A Palma, J A Jiménez-Tejada, J Banqueri and J A López-Villanueva, "Influence of Mobility Fluctuations on Random Telegraph Signal Amplitude in n-channel metal-oxide-semicondutor field-effect transistors", *J.Appl.Phys.*, vol. 82, , pp. 4621-4628, Nov. 1997.

[6] P. Stolk, F P Widdershoven, D B M Klaassen, "Modeling Statistical Dopant Fluctuations in MOS Transistors", *IEEE Trans. El. Dev.*, vol. 45, pp. 1960-1971, Sept. 1998.

[7] U. Schaper, C G Linnenbank, amd R Thewes, "Precise characterization of long-distance mismatch of CMOS devices", *IEEE Trans. Sem. Man.*, vol. 14, pp. 311-317, Nov. 2001.

[8] G. Wirth, U Hilleringmann, J T Horstmann and K Goser, "Negative Differential Resistance in Ultrashort Bulk MOSFETS", in *Annual Conference of the IEEE Industrial Electronics Society (Proceedings)*, 1999, pp. 29-34.

[9] MiniMos Manual "http://www.iue.tuwien.ac.at/", 2003

[10] P Dutta and P M Horn, "Low-Frequency Fluctuations in Solids: 1/f Noise", *Reviews of Modern Physics*, vol. 53, p. 497-516, July 1981.

[11] M J Kirton and M J Uren, "Noise in Solid-State Microstructures: A New Perspective on Individual Defects, Interface States and Low-Frequency (1/f) Noise", *Advances in Physics*, vol. 38, p. 367-468, 1989.

[12] S. Christensson, I. Lundström, and C. Svensson, "Low Frequency Noise in MOS Transistors – I Theory", *Solid-St. El.*, vol. 11, pp. 791-812, 1968.

[13] G. Reimbold, "Modified 1/f Trapping Noise Theory and Experiments in MOS Transistors from Weak to Strong Inversion-Influence of Interface States", *IEEE Trans. El. Dev.*, vol. 31, pp. 1190-1998, Sept. 1984.

[14] S.M. Sze, *Physics of semiconductor devices*, 2$^{nd}$ ed. New York: Wiley , 1981

[15] BSIM4 Manual "http://www-device.EECS.Berkeley.EDU/~bsim4/", 2001

[16] S. Villa, A. Lacaita, L. Perron and R. Bez, "A physically-Based Model of the Effective Mobility in Heavily-Doped n-MOSFETs", *IEEE Trans. El. Dev.*, vol. 45, pp. 110-115, Jan. 1998.

[17] S.T. Martin, G.P. Li, and J. Worley, "The Gate Bias and Geometry Dependence of Random Telegraph Signal Amplitudes", *IEEE Electron Device Letters*, vol. 18, pp. 444-446, Sept. 1997.

[18] R. Brederlow, PhD thesis, TU Berlin, 1999.

[19] K.K. Hung, P.K. Ko, C. Hu, and Y.C. Cheng, "A Unified Model for the Flicker Noise in Metal-Oxide-Semiconductor Field-Effect Transistors", *IEEE Trans. El. Dev.*, vol. 37, pp. 654-665, March 1990.

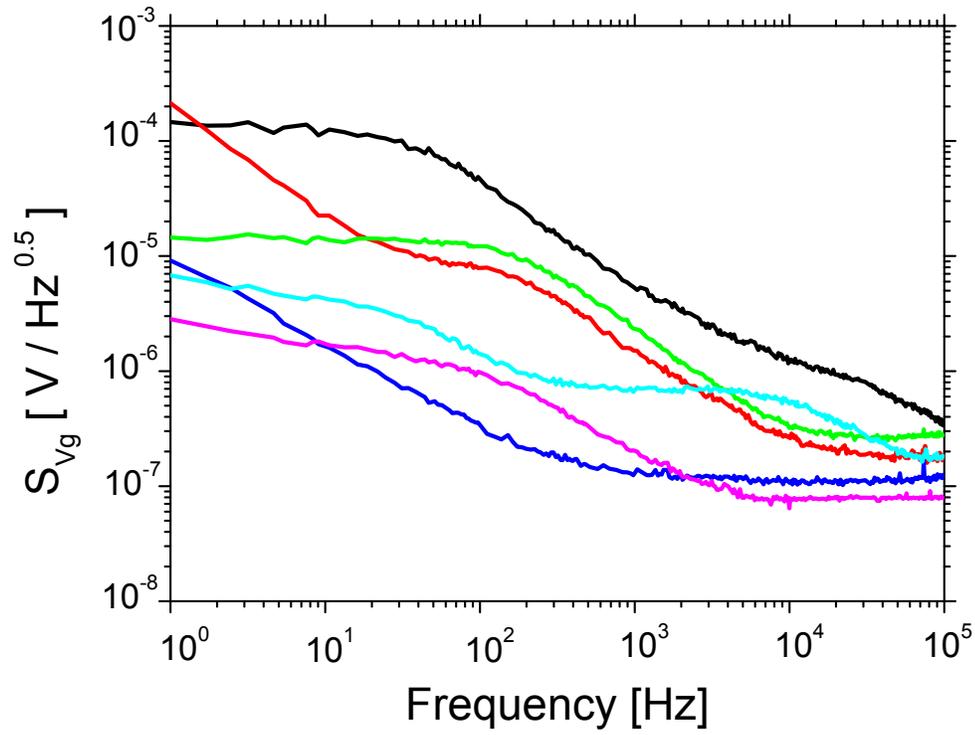

Figure 1: Gate referred voltage noise of 6 different W=0.16μm/L=0.13μm n-MOS transistors from a 0.13μm standard CMOS process with $t_{ox}$=2.2nm and $V_{th}$ = 300 mV. Characterization in saturation at $V_G$ = 0.55V and $V_D$ = 1V.

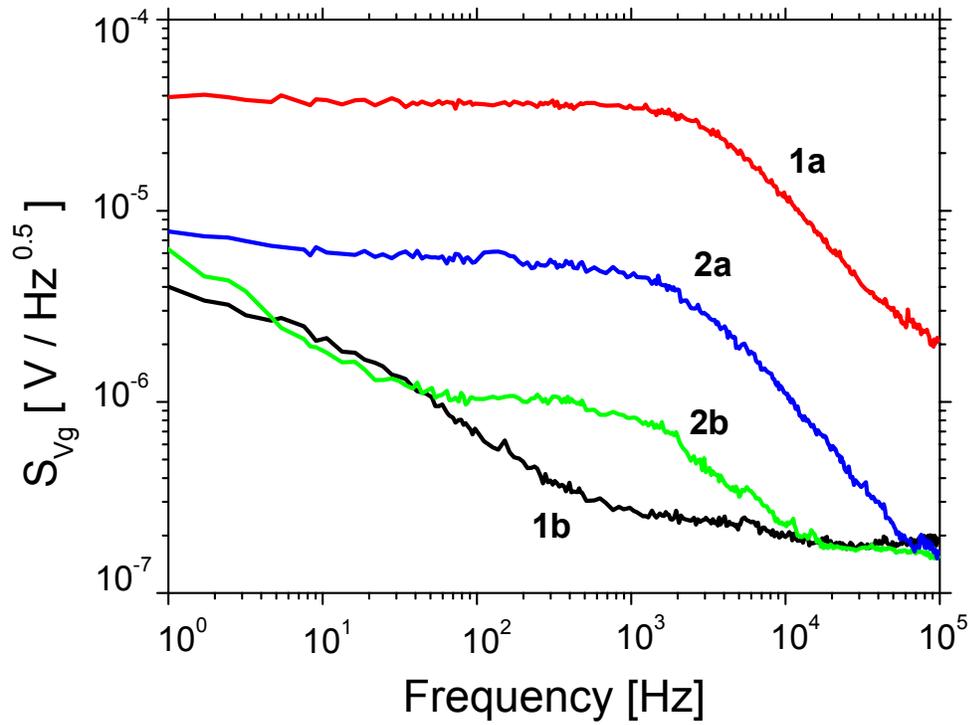

Figure 2: Gate referred voltage noise of two different W=0.16µm/L=0.13µm n-MOS transistors from a 0.13µm standard CMOS process with $t_{ox}$=2.2nm and $V_{th}$=300mV under different characterization conditions. Curves (1a) and (1b): first device biased at $V_G$=0.85V, with $V_D$=0.15V and $V_D$=1.0V, respectively; curves (2a) and (2b): second device, biased at $V_D$=1.0V, with $V_G$=0.85V and $V_G$=0.55V, respectively.

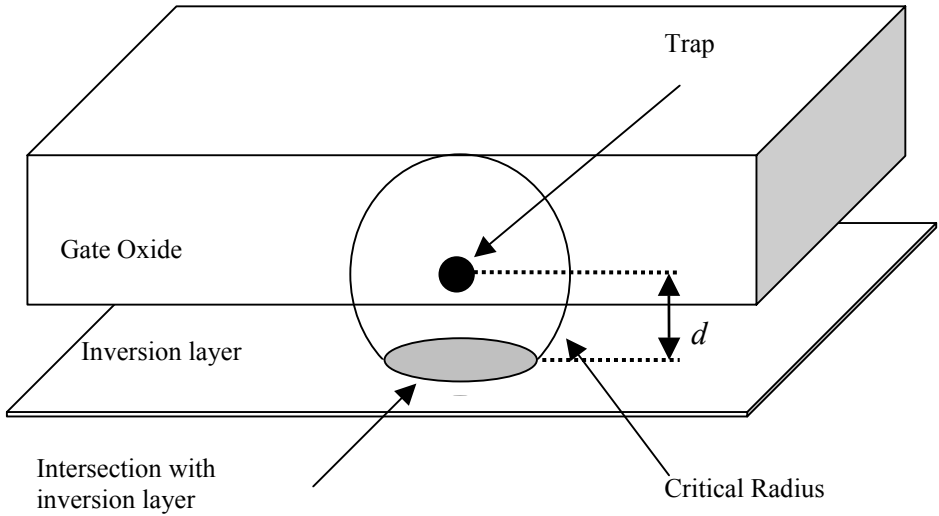

Figure 3: Schematic plot of the inversion layer of a MOS transistor disturbed by an occupied trap state. From elementary geometry considerations, the radius $r_i$ of the intersection between the channel plane (inversion layer) and the sphere defined by the critical radius $r_c$ is calculated according to eq. (34), for $d < r_c$. If $d \geq r_c$, it is assumed that the trap causes no scattering ($\delta\mu = 0$).

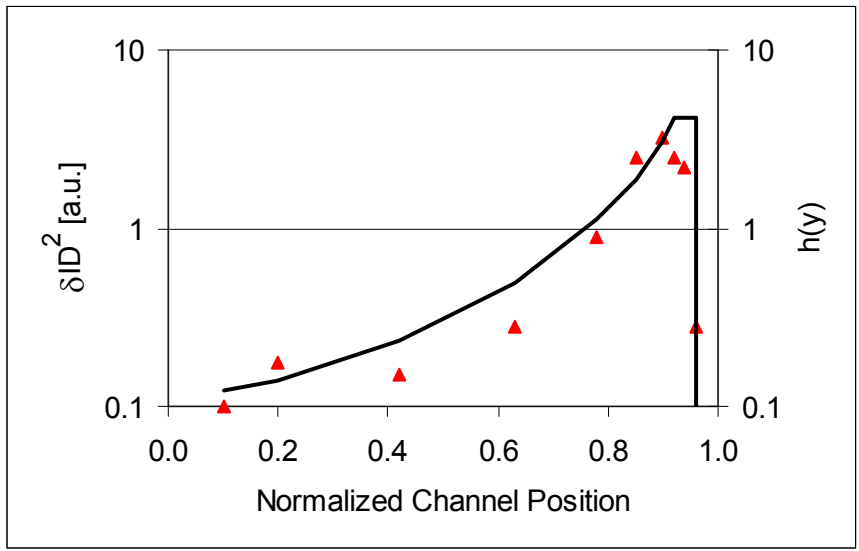

Figure 4: Triangles: Simulated contribution from traps at different channel positions (source at $x=0$ and drain at $x=1$) on the trap related noise of a MOSFET operated in saturation (left axis). Full line: efficiency term $h(y)$ as evaluated from eq. (38) (right axis).

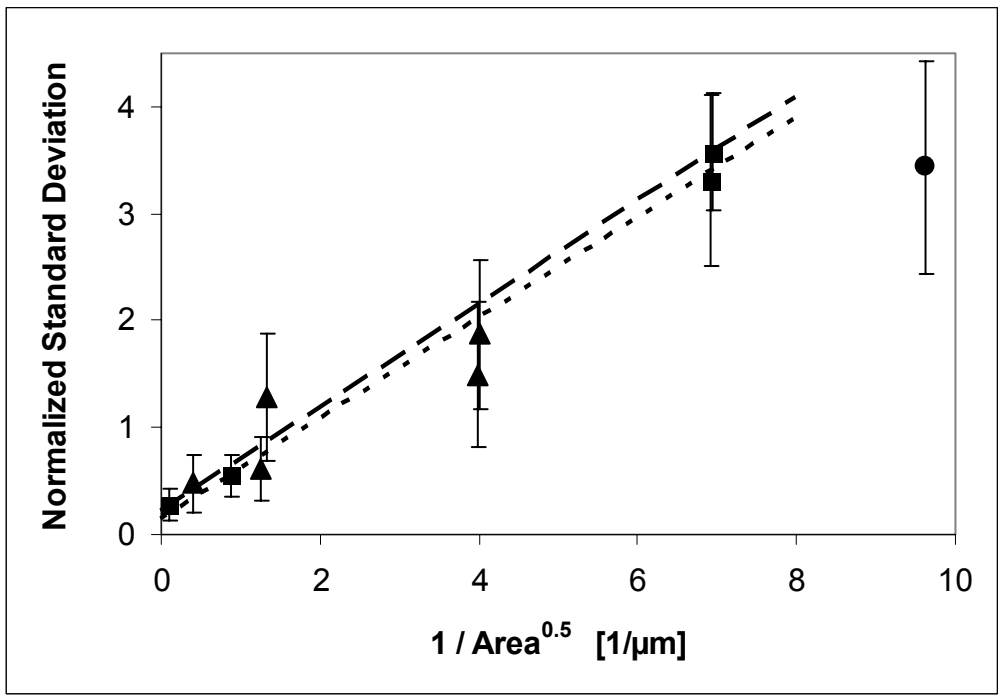

Figure 5: Normalized standard deviation of gate referred voltage noise in the bandwidth 1 Hz to 10 kHz versus area$^{-0.5}$ for transistors biased in saturation. Error bars are $2\sigma$-values of the measurement accuracy.

▲ 0.25μm technology node ($L_{min}$=0.25μm, $t_{ox}$=5nm). Total of 30 transistors measured.

■ 0.13μm technology node ($L_{min}$=0.13μm, $t_{ox}$=2.2nm). Total of 127 transistors measured.

● 90nm technology node ($L_{min}$=0.09μm, $t_{ox}$=1.6nm). Total of 14 transistors measured.

The dashed line shows results for the 0.13μm node calculated on the basis of eq. (49). Dotted line shows results for the 0.25μm node. (Normalized standard deviation is standard deviation of the square of gate referred voltage noise *divided by* the average of the square of gate referred voltage noise.)

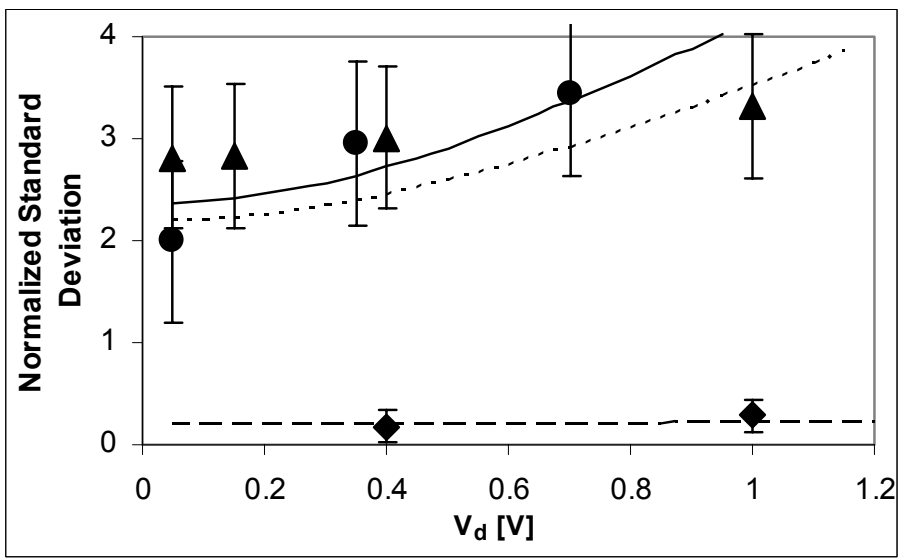

Figure 6: Normalized standard deviation of the gate referred voltage noise versus drain voltage for W/L=0.16μm/0.12μm p-MOS transistors operated at $V_g$ - $V_{th}$ = 0.25V (▲), W/L=10μm/10μm n-MOS transistors at $V_g$ - $V_{th}$ = 0.25V (♦) and W/L=0.12μm/0.09μm n-MOS transistors at $V_g$ - $V_{th}$ = 0.15V (●). The full line shows the result of a calculation based on the model equations for W/L=0.12μm/0.09μm n-MOS transistors (90nm technology – all other data from 0.12μm technology), the dashed line for the W/L=10μm/10μm n-MOS transistors, and dotted line for the W/L=0.16μm/0.12μm p-MOS transistors.

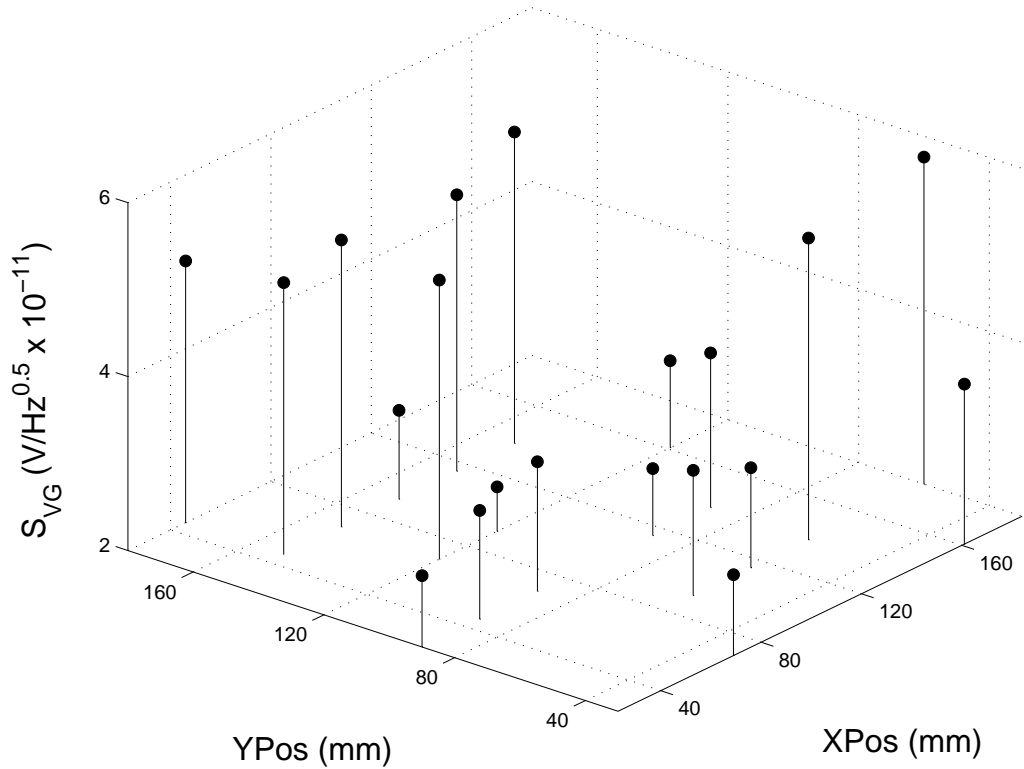

Figure 7: Gate referred voltage noise S_{VG} (bandwidth 1Hz-100kHz) of 20 n-MOSFETs from a 0.13µm standard CMOS process with t_{ox}=2.2nm, at different wafer positions. Device dimensions are W=L=10µm, characterization is performed in saturation at Vg-V_{th}=0.25V and V_{d}=1.0V. X and Y axis indicate die *x* and *y* position in millimeters, on a 200mm wafer. The measurements have been performed on a test waver. Therefore not all positions show target device data. These positions are not included here.

Table 1: Average number of traps per transistor $<N_{tr}>$, $1/N_{dec}$ and $<A^4>/<A^2>^2$. Data is from minimum area transistors of each technology node.

| Technology Node | $W \times L$ (µm×µm) | $\sigma_{Svg}$ Normalized | $<N_{tr}>$ | $(N_{dec} * WL)^{-0.5}$ | $(<A^4>/<A^2>^2)^{0.5}$ |
|---|---|---|---|---|---|
| 0.25µm | 0.30x0.25 | 1.87 | 9[*] | 1.01 | 1.85 |
| 0.13µm | 0.16x0.13 | 3.57 | 2.7[+] | 1.85 | 1.93 |
| 0.09µm | 0.12x0.09 | 3.43 | 1.8[+] | 2.26 | 1.52 |
| * Extracted from charge pumping measurements and LF-noise data. | | | | | |
| + Extracted from LF-Noise data. | | | | | |